 \documentclass[11pt,floatfix,notitlepage]{article}
 \usepackage[english]{babel}
 \usepackage{amsmath,mathrsfs,amssymb,esint,cancel}
 \usepackage{pstricks,pst-plot,pst-grad,pst-3dplot,pstricks-add,pst-node}
 \usepackage{pst-eps,auto-pst-pdf}
 \usepackage{caption2,graphics,placeins,verbatim}
 \usepackage[totalwidth=500pt, totalheight=650pt]{geometry}
 \usepackage{etex}
 \usepackage{extarrows}
 \usepackage{CJKutf8}
 \usepackage[unicode]{hyperref}
 \usepackage{indentfirst}
 \usepackage{setspace}
 \newcommand{\beq}[1]{\begin{equation}\label{#1}}
 \newcommand{\eeq}{\end{equation}}
 \newcommand{\bea}[1]{\begin{eqnarray}\label{#1}}
 \newcommand{\eea}{\end{eqnarray}}
 \newcommand{\bfk}{\mathbf{k}}
 \newcommand{\bfx}{\mathbf{x}}
 \newcommand{\ch}{\mathrm{ch}}
 \newcommand{\sh}{\mathrm{sh}}
  
 \newcommand{\decimalA}[1]{{
 \psset{unit=#1}%
 \begin{pspicture}(-3,-3)(3,3)
 \multido{\ix=-2+1}{5}{\multido{\iy=-2+1}{5}{
 \ifodd\numexpr(\ix+\iy)
 \psdot[dotstyle=x,dotscale=1,linecolor=blue](\ix,\iy)
 \else
 \psdot[dotstyle=*,dotscale=1,linecolor=red](\ix,\iy)
 \fi
 \rput(\ix,\iy){\psline(-0.5,0)(-0.2,0)\psline(0.2,0)(0.5,0)
 \psline(0,-0.5)(0,-0.2)\psline(0,0.2)(0,0.5)}
 }}
 \end{pspicture}%
 }}
 \newcommand{\decimalB}[1]{{
 \psset{unit=#1}%
 \begin{pspicture}(-3,-3)(3,3)
 \multido{\ix=-2+1}{5}{\multido{\iy=-2+1}{5}{
 \ifodd\numexpr(\ix+\iy+1)
 \psdot[dotstyle=*,dotscale=1,linecolor=red](\ix,\iy)
 \rput(\ix,\iy){\psline(-0.5,0.5)(-0.2,0.2)\psline(0.2,-0.2)(0.5,-0.5)
 \psline(0.2,0.2)(0.5,0.5)\psline(-0.5,-0.5)(-0.2,-0.2)}
 \fi
 }}
 \end{pspicture}%
 }}
 \newcommand{\decimalC}[1]{{
 \psset{unit=#1}%
 \begin{pspicture}(-3,-3)(3,3)
 \multido{\ix=-2+1}{5}{\multido{\iy=-2+1}{5}{
 \ifodd\numexpr(\ix+\iy+1)
 \ifodd\numexpr(\ix+1)
 \psdot[dotstyle=*,dotscale=1,linecolor=red](\ix,\iy)
 \else
 \psdot[dotstyle=x,dotscale=1,linecolor=blue](\ix,\iy)
 \fi
 \rput(\ix,\iy){\psline(-0.5,0.5)(-0.25,0.25)\psline(0.25,-0.25)(0.5,-0.5)
 \psline(0.25,0.25)(0.5,0.5)\psline(-0.5,-0.5)(-0.25,-0.25)}
 \fi
 }}\end{pspicture}%
 }}
 \newcommand{\decimalD}[1]{{
 \psset{unit=#1}%
 \begin{pspicture}(-3,-3)(3,3)
 \multido{\ix=-2+1}{5}{\multido{\iy=-2+1}{5}{
 \ifodd\numexpr(\ix+\iy+1)
 \ifodd\numexpr(\ix+1)
 \psdot[dotstyle=*,dotscale=1,linecolor=red](\ix,\iy)
 \rput(\ix,\iy){\psline(-1,0)(-0.4,0)\psline(0.4,0)(1,0)
 \psline(0,0.4)(0,1)\psline(0,-0.4)(0,-1)}
 \fi
 \fi
 }}\end{pspicture}%
 }}
 \newcommand{\decimalE}[1]{{
 \psset{unit=#1}%
 \begin{pspicture}(-2,-2)(2,2)
 \multido{\ix=-1+1}{3}{\multido{\iy=-1+1}{3}{
 \ifodd\numexpr(\ix+\iy+1)
 \psdot[dotstyle=*,dotscale=1,linecolor=red](\ix,\iy)
 \else
 \psdot[dotstyle=x,dotscale=1,linecolor=blue](\ix,\iy)
 \fi
 \rput(\ix,\iy){\psline(-0.5,0)(-0.2,0)\psline(0.2,0)(0.5,0)
 \psline(0,-0.5)(0,-0.2)\psline(0,0.2)(0,0.5)}
 }}\end{pspicture}%
 }}
 
 \newrgbcolor{ircolor}{0.95 0.8 0.7}
 \newrgbcolor{uvcolor}{0.1 0.1 0.8}
 \newcommand{\dOneRGflow}[1]{{\psset{unit=#1}
 \begin{pspicture}(-0.5,-0.5)(4.2,4.2)\psframe(-0.5,-0.5)(4.2,4.2)
 \psaxes[ticks=none,arrows=->](0,0)(-0.5,-0.5)(3.8,3.6)
 \pscustom[linestyle=none, fillstyle=gradient,gradangle=-45,gradbegin=blue,gradend=red,gradmidpoint=0]{
 \parametricplot[algebraic=true]{0}{1.7}{2*t|2*t}
 \parametricplot[algebraic=true]{1.9}{0}{2*t|ln(0.5*(2.71828^(2*t)+2.71828^(-2*t)))}
 }
 {\psset{linestyle=dashed,dash=0.6pt 0.6pt}
 \psline{->}(3.2, 3.2)(3.2, 2.50851)
 \psline{->}(3.2, 2.50851)(2.50851, 2.50851)(2.50851, 1.82197)
 \psline{->}(2.50851, 1.82197)(1.82197, 1.82197)(1.82197, 1.15463)
 \psline{->}(1.82197, 1.15463)(1.15463, 1.15463)(1.15463, 0.556191)
 \psline{->}(1.15463, 0.556191)(0.556191, 0.556191)(0.556191, 0.147301)}
 \rput(4,0){$k$}
 \rput(0,3.8){$k'$}
 \end{pspicture}
 }}
 \newcommand{\dTwoRGflow}[1]{{\psset{unit=#1}
 \begin{pspicture}(-0.5,-0.5)(4.2,4.2)\psframe(-0.5,-0.5)(4.2,4.2)
 \psaxes[ticks=none,arrows=->](0,0)(-0.5,-0.5)(3.8,3.6)
 \pscustom [linestyle=none, fillstyle=gradient,gradangle=-45,gradbegin=blue,gradend=red,gradmidpoint=0]{\parametricplot[algebraic=true]{0}{0.6094}{4*t|4*t}
 \parametricplot[algebraic=true]{0.6094}{0}{4*t|4*ln(0.5*(2.71828^(2*t)+2.71828^(-2*t)))}
 }\pscustom [linestyle=none, fillstyle=gradient,gradangle=135,gradbegin=blue,gradend=red,gradmidpoint=0]{\parametricplot[algebraic=true]{0.6094}{0.9}{4*t|4*t}
 \parametricplot[algebraic=true]{0.8}{0.6094}{4*t|4*ln(0.5*(2.71828^(2*t)+2.71828^(-2*t)))}
 }
 \psdot[dotstyle=|,dotscale=2](2.6,0)\rput(2.6,-0.3){$k_c$}
 {\psset{linestyle=dashed,dash=0.6pt 0.6pt}
 \psline(2.3, 2.3)(2.3, 2.20959)(2.20959, 2.20959)
 (2.20959, 2.06312)
 \psline{->}(2.20959, 2.06312)(2.06312, 2.06312)(2.06312, 1.83209)
 \psline{->}(2.06312, 1.83209)(1.83209, 1.83209)(1.83209, 1.48554)
 \psline{->}(1.83209, 1.48554)(1.48554, 1.48554)(1.48554, 1.01476)
 \psline{->}(1.48554, 1.01476)(1.01476, 1.01476)(1.01476, 0.494186)
 \psline{->}(1.01476, 0.494186)(0.494186, 0.494186)(0.494186, 0.120887)
 \psline(2.58, 2.58)(2.58, 2.67957)(2.67957, 2.67957)(2.67957, 2.85193)
 \psline{->}(2.67957, 2.85193)(2.85193, 2.85193)(2.85193, 3.15578)
  \psline{->}(2.85193, 3.15578)(3.15578, 3.15578)(3.15578, 3.70585)
  }
 \rput(4,0){$k$}
 \rput(0,3.8){$k'$}
 \end{pspicture}
 }}
 \newcommand{\desiredRGflow}[1]{{\psset{unit=#1}
 \begin{pspicture}(-0.5,-0.5)(4.2,4.2)\psframe(-0.5,-0.5)(4.2,4.2)
 \psaxes[ticks=none,arrows=->](0,0)(-0.5,-0.5)(3.8,3.6)
 \pscustom [linestyle=none, fillstyle=gradient,gradangle=-55,gradbegin=blue,gradend=red,gradmidpoint=0]{
 \parametricplot[algebraic=true]{0}{0.6094}{4*t|4*t}
 \parametricplot[algebraic=true]{0.6094}{0}{4*t|2*ln(2.71828^t+sqrt(2.71828^(2*t)-1))}
 }\pscustom [linestyle=none, fillstyle=gradient,gradangle=125,gradbegin=blue,gradend=red,gradmidpoint=0]{
 \parametricplot[algebraic=true]{0.6094}{0.86}{4*t|4*t}
 \parametricplot[algebraic=true]{0.9}{0.6094}{4*t|2*ln(2.71828^t+sqrt(2.71828^(2*t)-1))}
 }
 \psdot[dotstyle=|,dotscale=2](2.6,0)\rput(2.6,-0.3){$k_c$}
 {\psset{linestyle=dashed,dash=0.6pt 0.6pt}
 \psline{->}(0.32, 0.32)(0.32, 0.810708)
 \psline{->}(0.32, 0.810708)(0.810708, 0.810708)(0.810708, 1.31677)
  \psline{->}(0.810708, 1.31677)(1.31677, 1.31677)(1.31677, 1.71314)
  \psline{->}(1.31677, 1.71314)(1.71314, 1.71314)(1.71314, 1.98554)
  \psline{->}(1.71314, 1.98554)(1.98554, 1.98554)(1.98554, 2.16097)
  \psline(1.98554, 2.16097)(2.16097, 2.16097)(2.16097, 2.27018)(2.27018, 
  2.27018)(2.27018, 2.33691)
  \psline{->}(3.32, 3.32)(3.32, 2.94363)
  \psline{->}(3.32, 2.94363)(2.94363, 2.94363)(2.94363, 2.73199)
  \psline(2.94363, 2.73199)(2.73199, 2.73199)(2.73199, 2.61044)(2.61044, 2.61044)(2.61044, 2.53964)(2.53964, 2.53964)
  }
 \rput(4,0){$k$}
 \rput(0,3.8){$k'$}
 \end{pspicture}
 }}
 
 \newrgbcolor{magenta}{0.5 0 0.5}
 \newrgbcolor{darkgreen}{0 0.4 0.6}
 \newcommand{\chainRenormalize}[1]{{\psset{unit=#1}{
 \begin{pspicture}(-4,-3.5)(2.8,0.5)
 \psline{->}(-2.3,0.2)(-2.3,-3.4)
 \rput(-1.8,-0.1){\blue$\scriptstyle N=0$}
 \rput(-1.9,-3){\red$\scriptstyle\infty$}
 {\psset{linecolor=blue}
 \rput(-3.1,-0.1){\blue$\scriptstyle kss'\!+\!k_0$}
 \multido{\rx=-2.4+0.6}{9}{
 \psline[dotstyle=|](\rx,-0.1)(\rx,0.1)
 \rput(\rx,0.15){\psarc[linewidth=0.1pt]{<-}(0.13,0){0.13}{0}{180}
 \psarc[linewidth=0.1pt]{->}(-0.13,0){0.13}{0}{180}
 }}}
 {\psset{linecolor=green}
 \rput(-3.1,-0.9){\green$\scriptstyle kss'\!+\!k_0$}
 \multido{\rx=-2.1+0.6}{8}{\psline[dotstyle=|](\rx,-0.1)(\rx,0.1)}
 \multido{\rx=-1.8+1.2}{4}{
 \psline[dotstyle=|](\rx,-1.1)(\rx,-0.9)
 \rput(\rx,-0.85){\psarc[linewidth=0.2pt]{<-}(0.25,0){0.25}{0}{180}
 \psarc[linewidth=0.2pt]{->}(-0.25,0){0.25}{0}{180}
 }}}
 {\psset{linecolor=red}
 \rput(-3.1,-1.9){\yellow$\scriptstyle kss'\!+\!k_0$}
 \multido{\rx=-1.2+1.2}{3}{\psline[dotstyle=|, linewidth=2pt](\rx,-1.1)(\rx,-.9)}
 \multido{\rx=-1.2+2.4}{2}{
 \psline[dotstyle=|, linewidth=2pt](\rx,-2.1)(\rx,-1.9)
 }\pnode(-1.2,-1.8){A}\pnode(1.2,-1.8){B}\pnode(0,-1.8){O}
 \nccurve[angleA=110,angleB=60,linewidth=0.5pt]{->}{O}{A}
 \nccurve[angleA=60,angleB=120,linewidth=0.5pt]{->}{O}{B}
 }
 {\psset{linecolor=orange}
 \rput(-3.1,-2.9){\red$\scriptstyle kss'\!+\!k_0$}
 \psline[dotstyle=|, linewidth=2pt](0,-2.1)(0,-1.9)
 \multido{\rx=-1.2+2.4}{2}{
 \psline[dotstyle=|, linewidth=2pt](\rx,-3.1)(\rx,-2.9)
 }}
 \end{pspicture}%
 }}}
  \newcommand{\adsTwo}[1]{{
 \psset{unit=#1}
 \begin{pspicture}(-2.3,-3.5)(3,0.5)
 \parametricplot[algebraic=true]{-3.3}{0.2}{0.25+1.5*2^t|-3-t}
 \parametricplot[algebraic=true]{-3.3}{0.2}{-0.25-1.5*2^t|-3-t}
 \psline[linewidth=0.5pt](-0.402,0.3)(0.402,0.3)
 \psline[linewidth=0.2pt](-1.98,-3.2)(1.98,-3.2)
 \psline{->}(-2,0.2)(-2,-3.4)
 \rput(-1.7,0.2){\blue$\scriptstyle\tau=0$}
 \rput(2.2,0.2){\blue$\scriptstyle\frac{-d\tau^2+d\vec{\mathbf{x}}^2}{f^2(\tau)}$}
 \rput(-1.8,-3.3){\red$\scriptstyle\infty$}
 \rput(2.2,-3.1){\red$\scriptstyle\frac{-d\tau^2+d\vec{\mathbf{x}}^2}{\tau^2/\ell^2}$}
 \psdots(-0.6,-2.5)(0.6,-2.5)
 \rput(-0.85,-2.5){$A$}\rput(0.85,-2.5){$C$}
 \parametricplot[algebraic=true]{-0.6}{0.6}{t|-3*t^2+1.08-2.5}
 \parametricplot[algebraic=true,linestyle=dashed]{-0.6}{0.6}{t|t^2-0.36-2.5}
 \end{pspicture} 
 }}
 
 \savedata{\atholo}[{{0., 0.0003}, {0.0101632, 0.17444}, {0.0640704, 0.296191}, {0.169646,
   0.397654}, {0.313745, 0.494313}, {0.479393, 0.597472}, {0.653743, 
  0.714196}, {0.828855, 0.848205}, {1.00027, 1.00122}, {1.16568, 
  1.17399}, {1.32401, 1.36682}, {1.47491, 1.57984}, {1.61843, 
  1.81309}, {1.75485, 2.06661}, {1.88453, 2.34039}, {2.0079, 2.63445}}]
  \savedata{\atlcdm}[{{0.04, 0.0224689}, {0.138, 0.191974}, {0.236, 0.302205}, {0.334, 
  0.397556}, {0.432, 0.486181}, {0.53, 0.571873}, {0.628, 
  0.656954}, {0.726, 0.743087}, {0.824, 0.831591}, {0.922, 
  0.923593}, {1.02, 1.02011}, {1.118, 1.1221}, {1.216, 
  1.23049}, {1.314, 1.3462}, {1.412, 1.47017}, {1.51, 
  1.60336}, {1.608, 1.74677}, {1.706, 1.90146}, {1.804, 
  2.06853}, {1.902, 2.24919}, {2., 2.44469}}]
  \newcommand{\atCosmo}[2]{{
 \psset{xunit=#1,yunit=#2}
 \begin{pspicture}(-1,-1)(3,3.5)
 \dataplot[plotstyle=curve,linecolor=red]{\atholo}
 \dataplot[plotstyle=curve,linecolor=blue]{\atlcdm}
 \psaxes[Dx=0.5,Dy=0.5,tickstyle=top]{->}(-0.01,0)(-0.3,-0.3)(2.1,2.7)
 \rput(2.3,0){$H_0t$}
 \rput(0.2,2.5){$a(t)$}
 \end{pspicture}
 }}
 
\begin{document}
\begin{CJK}{UTF8}{gbsn}
\CJKtilde

\begin{center}
{\Large\bfseries Emergent time axis from statistic/gravity dualities}

Ding-fang Zeng

Applied Math and Physics School, Beijing University of Technology
\\
People's Republic of China, Bejing 100124
\end{center}

\abstract{We discuss a very naive but natural idea that time emerges as the holographic dimension of gauge systems in euclidean space, which take statistic, e.g. Ising model as concrete implementations. By identifying the renormalization group flow of statistic models with the time flow of dual gravities, we get a universe whose evolution history is qualitatively the same as our real world. We comment highlights projected by this idea on the cosmological constant problem and develope preliminary evidences for the validity of this idea.}

\tableofcontents

\section{Basic ideals on emergent space-time}
The gauge/gravity duality and cosmology are  two most fruitful branches of new century fundamental physics. Development of the latter \cite{CosmicObservation} brings us cosmological constant problem (CCP)--- the biggest problem \cite{CCproblem} of fundamental physics, while establishments of the former \cite{AdSCFT1998, AdSCFTreview} provide us powerful methods to solve questions appearing in various fields \cite{holoQCDreview,holoCMTreview,holoIntegralReview,holoNEquilibriumReview} of theoretical physics, including of course CCP \cite{Papadodimas2011, JKM2007}. 

Among the various ideas initiated by gauge/gravity dualies, the most attractive one may be the idea that the space-itself, may be emergent objects from the lower dimensional gauge theories \cite{LLM2004emst, ReyHikida2005emst, Berenstein2005emst, KM2009emst, papa2011emst, verlinde2011,swingle2012,qixiaoliang2013}. Taking the most famous anti-de Sitter space/conformal field theory correspondence (AdS/CFT) as an example
\beq{}
\mathrm{gravities~on~}ds^2=\frac{\ell^2}{z^2}(-dt^2+d\vec{\bfx}^2+dz^2)
\Leftrightarrow~\mathrm{CFT~on~}ds^2=-dt^2+d\vec{\bfx}^2
\label{AdScft}
\eeq
Since gavities on the left hand side of this duality is equivilantly described by field theories on the right lower-dimensional space-time and vice versa, peoples say that the dimension $z$ on the gravity side is emergent from lower dimensional gauge system. Observation and reflections on this duality \eqref{AdScft} naturally lead to questions, why must it be space, why not it be time that emerges from lower dimensional physics systems? For example, can the the following duality
\beq{}
\mathrm{gravities~on~}ds^2=\frac{\ell^2}{t^2}(-dt^2+d\vec{\bfx}^2)\Leftrightarrow
\mathrm{CFT~on~}ds^2=d\vec{\bfx}^2
\label{dScfte}
\eeq
be implemented physically thus let time emerge from physical systems living on flat euclidean spaces? We will call this duality as dS/CFTe(e is for euclidean, dS is for the left geometry of de Sitter in poincare coordinates). From symmetry matching aspects, this is very natural expectation since both sides of it have $SO(1,n+1)$ as the key symmetries.

In comparison to AdS/CFT's brane-dynamical constructuring \cite{AdSCFT1998,AdSCFTreview} and prosperous developing \cite{holoQCDreview,holoCMTreview,holoIntegralReview,holoNEquilibriumReview}, dS/CFT \cite{Strominger2001, KOP2002, Anninos2012, dScft2001Strominger, dScft2002Klemm, dSmatrix2002limiao, MRR2011, ParikhHoloDs2004, dSentropyMS1998, linFL1999, hawking2001} especially dS/CFTe has never been implemented/invented top-downly and has been only rarely(relative to AdS/CFT) explored in string communities due to the difficulty of de Sitter vaccum's brane-constructuring \cite{hullDs2001,kkltDs2003, bmDs2003}. However, in this paper we will provide a bottom-up way to constructure the following more general statistic/gravity dualities
\beq{}
\mathrm{gravities~on~}ds^2=\frac{-dt^2+d\vec{\bfx}^2}{f^2(t)}
\Leftrightarrow\mathrm{statistic~models~on~}ds^2=d\vec{\bfx}^2.
\label{statisGravity}
\eeq
where statistic models on the right hand side are just proxies of physic systems involving no time evolutions. Time flow in the left hand side will be identified with the renormalization group flow \cite{Migdal, Kadanoff} of the statistic models thus make time emerge naturally. We will use Ising model on descrete lattices as the concrete implementation of statistic models on euclidean space, see \cite{ising3Dexact} for recent devolopment on this model, \cite{Maloney2012} for its gravity dual aspects, and \cite{Skenderis2010, MaldacenaThreePoint, FRWCFTsusskind} for related holographic universe explorations. Of course, as a bottom-up construction, we have no any prior reason to limit the left hand side in conventional gravities and exclude other possibility such as higher spin theories \cite{vasilev1996HS, AdSONduality, HSyinxi2012, HSleigh2014}. 

We will give in the next section a short introduction to the concepts of renormalization and renormaization group flows, taking Ising model as the basic illustration. In the next next section, by identifing the renormalization group flow (RGF) of ising model with the inverse time arrow of cosmology, we constructure a universe whose evolution history qualitatively coincides with our real world and comment the highlights this idea might project on the solving of CCP. The next next next section contain our preliminary development of evidences for the statistic/gravity duality. The last section is our conclusion and discussion.

\section{Statistic model and renormalization group flow}
Ising model may be the simplest model to illustrate the basic ideal of statistic mechanic and field theories. The models consists of spins living on some periodic lattices. Each spin has only two possible states, up or down, i.e. $s_i=+1/-1$. All nearest neighboring spin-pairs interact through hamiltions $-ks_is_{i'}$. In finite temperatures, properties of the system is completely determined by the partition function
\beq{}
Z[k,k_0,\{h_i\}]=\sum_\mathrm{config}e^{-\beta H+\sum h_is_i},~\beta H=-\frac{1}{k_{_B}T}\sum_{\langle ii'\rangle}(\mu s_is_{i'}+\mu_0)\equiv\sum_{\langle ii'\rangle}(ks_is_{i'}+k_0)
\label{isingPartition}
\eeq
If we take zero length limit of the lattice interval and define the average of spins in each macro-large but micro-small volume elements as a continous field $\phi$, then the partition function will very naturally become the parth integration of $\phi$ field in euclidean spaces
\beq{}
Z[k,k_0,h]=\int\!\!\mathcal{D}\phi\,e^{\int\!d^n\!x\,[(\partial_i\phi)^2+m^2(k)\phi^2+\lambda(k)\phi^4+\cdots+\phi\cdot h]}
\eeq
This is just the reason we prefer to use Ising model as the concrete implementation of euclidean field theories in this paper.

\begin{figure*}[ht]
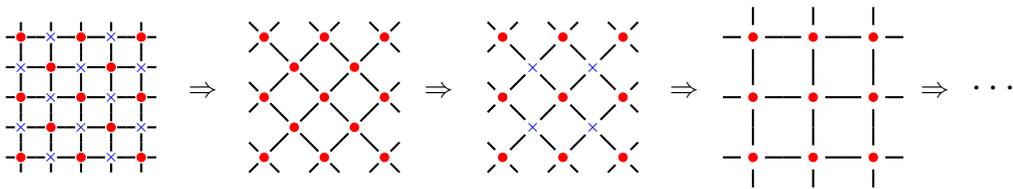

\begin{center}
\decimalA{0.4}
 ~\raisebox{12mm}{$\Rightarrow$}~
\decimalB{0.4}
\raisebox{12mm}{$\Rightarrow$}
\decimalC{0.4}
 ~\raisebox{12mm}{$\Rightarrow$}~~
\decimalD{0.4}
~\raisebox{12mm}{$\Rightarrow$}~~
~\raisebox{12mm}{\Large$\cdots$}~~
 \caption{The renormalization idea of ising models on 2-D squre lattices}
 \end{center}
 \label{figrenormalization}
 \end{figure*}

Either from the aspect of ising model or from that of effective field theories, renormalization group is all nothing but an alternative method of partition-function/path-integral calculation of the system. Taking ising model as examples, the relevant operations cosist of two self-repeating steps only. The first is dividing sites on the lattice into two types, say crossed and dotted, the latter of which forms a completely similar lattice as the parent one. The second is summing over the crossed-site states and attributing the summation effects to the redefinition of model parameters on the remaining lattice and going back the first step repeatedly. In mathematica formulaes, this is
\beq{}
\sum^\mathrm{config}_\mathrm{crossed~sites}\!\!\!e^{\sum ks_is_{i'}+k_0}
= e^{\sum k\!\!'~s_is_{i''}+k_0'}
\eeq
iterations of this operation leads to renormalization group naturally, see figure \ref{figrenormalization} for illustrations. 
For 2- and higher-dimensional ising models, the single-time renormalizing operation cannot be done exactly. But by the so called Migdal-Kadanoff approximation\cite{Migdal,Kadanoff}, we have
\beq{}
k'=\xi^{n-1}D^{-1}(\xi D(k)),~k_0'=\xi^{n-1}D_0^{-1}(\xi D(k_0,k),\xi D(k))
\label{Migdalformula}
\eeq
\beq{}
D(k)\equiv-\frac{1}{2}\ln(\mathrm{th}2k)=D^{-1}(k),~\xi=q+1
,~D_0(k_0,k)=k_0+\frac{1}{2}\ln2\mathrm{sh}2k
\nonumber
\eeq
where $n$ is the dimension of model lattices and $q$ is the number of sites every other which the state is summized on each dimension. In the figure \ref{figrenormalization} example, $n,q=2,1$ respectively. From equation \eqref{Migdalformula}, we can also prove that
\beq{}
\beta(k)\equiv\xi\frac{d k}{d\xi}|_{\xi=1}=(n-1)k-\frac{D(k)}{\mathrm{sh}(2k)}.
\label{betaOfk}
\eeq

\begin{figure*}[ht]
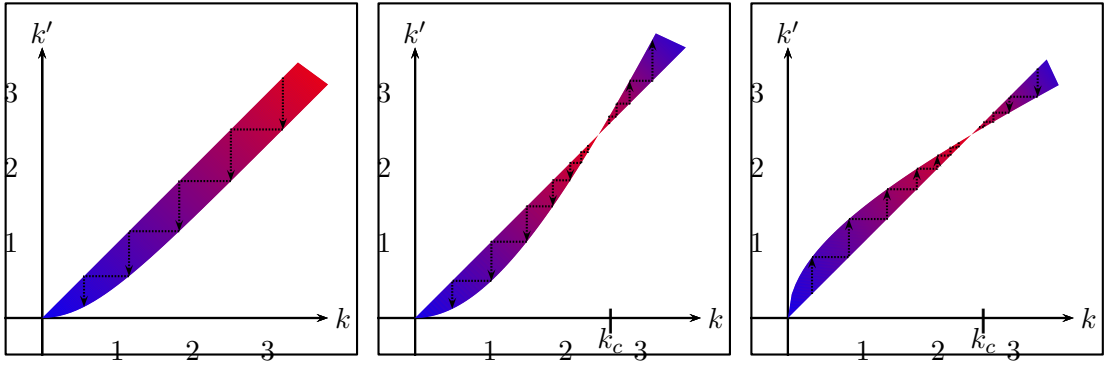

\setcaptionwidth{0.85\textwidth}
\begin{center}
\dOneRGflow{1}
\dTwoRGflow{1}
\desiredRGflow{1}
\caption{Renormalization group flows of 1-D, 2-D ising models with unstable infra-red  fixed point and some unknown statistic/field-theory models with desired stable fixed point.}
\label{figRGFlow}
\end{center}
\end{figure*}

The renormalizing transformation \eqref{Migdalformula} has two fixed points, $k=0$ and $k=k_c$(infinte in 1-D case), the latter of which corresponds to a non-trivial phase transition. Since on this point, the system developes long range orders and long range correllation, we call this point as infra-red fixed point. But for Ising model, this is a non-stable fixed point, in the sense that enough times of renormalization will bring the system far away from that point no matter how close the starting point is from it. For applications in the following section of this paper, we are more favor of statistic models with stable infra-red fixed point. The right most part of figure \ref{figRGFlow} gives such an example. Although currently we know no explicit realization of such models, we believe that such models exist in physics and may be related with Ising models very closely, since renormalization rules used in this example are just the inverse of that of Ising models. The final thing worthy of noticing is that at the beginning of renormalization group actions, the ising model parameter $k_0$ is an arbitrary additive constant which has no effects on statistical correlation functions, so we can set it to zero for simplicities. However, as renormalizations progress on, $k_0$ will become nonzero unavoidably. From effective field theories, this is just the feature of vacuum energies. 

\section{Emergent time axis from statistic models}
\begin{figure*}[t]
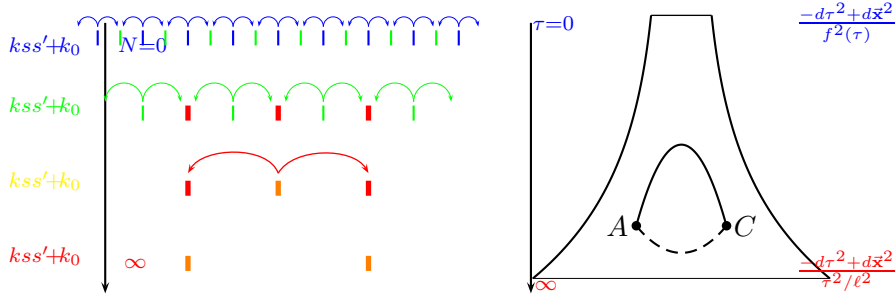

\setcaptionwidth{0.85\textwidth}
\begin{center}
\chainRenormalize{1}
\adsTwo{1}
\caption{The renormalization group flow of 1-D ising model and the geometry of (1+1)-D expanding universe. The $\tau$ coordinate in this figure can be obtained from the $\tilde{t}$ in \eqref{dualMetrics} or $t$ in \eqref{dualDesitter} through simple transformations. The geodesic line  connecting two equal-$\tau$ point A and C in the right hand part are bending upwards(solid line). In the dual Ising model, this corresponds to the fact that the shortest path connecting two nearby sites is through appropriate backward-running along the renormalization group flow. If we change the sign in front of the $d\tau^2$ term, the corresponding geodesic line will bend downwards(dash line), which is obviously inconsistant with the dual physic pictures.}
\label{figIsingdeSitter}
\end{center}
\end{figure*}

The flow of renormalization groups in equilabrium statistic system(ESS)/euclidean field theory(EFT) models is irreversable. This feature naturally reminds us the arrow of time in cosmologies. So let us make the key try of this paper to identify this two concepts. Referring to figure \ref{figIsingdeSitter}, by identifying the lattice of statistic models to the equal-time section of the universe, the order of renormalizations in statistic models to the time coordinate of the universe, the coupling constant of statistic models to the dilaton-factor of the dual geometry, we can write down the Einstein frame metric of the dual theory as follows,
\beq{}
ds^2=e^{2\phi(\tilde{t})}(-d\tilde{t}^2+e^{2\tilde{t}/\ell_0}d\vec{\bfx}^2),~e^{\phi(\tilde{t})}=k(\xi),~\xi=e^N=e^{\tilde{t}/\ell_0}
\label{dualMetrics}
\eeq
where $\ell_0$ is just constant with length dimension and $k(\xi)$ should be determined from renormalization group equations(RGE) like \eqref{betaOfk} but with a minus sign change (so that $\beta'(k_c)<0$) to implement stable infra-red fixed points. 
\beq{}
\beta(k)\equiv\xi\frac{dk}{d\xi}|_{\xi=1}=-(n-1)k+\frac{D(k)}{\mathrm{sh}(2k)}
\label{betaUniverse}
\eeq
The minus sign in front of the $d\tilde{t}^2$ term in \eqref{dualMetrics} is justified by geodesic analysis in the corresponding space-time, see captions in figure \ref{figIsingdeSitter}.

Near the infra-red fixed point of RGE \eqref{betaUniverse}, $\xi\rightarrow\infty$ and $k\approx k_c+\xi^{\beta'(k_c)}\rightarrow k_c$. 
For $n=1$ statistics, $k_c=\infty$. The dual (1+1)-D universe has future asymptotics
\beq{}
ds^2=-dt^2+a^2(t)dx^2,~a(t)\approx\frac{1}{4}\ln(32t/\ell_0)e^{(4t/\ell_0)/[\ln(32t/\ell_0)]}
\label{latterAsymptoticB}
\eeq
which if looked as an approximate de Sitter geometry will have infinite horizon size $\frac{\ell_0}{2}\ln(32t/\ell_0)]$ as $t\rightarrow\infty$.  For $n\geqslant2$ models, $k_c$ is finite. The dual geometry in the future limit $k\rightarrow k_c$ asymptotes to a pure de Sitter space-time
\beq{}
ds^2=-dt^2+e^{2t/(k_c\ell_0)}dx^2.
\label{dualDesitter}
\eeq
It's important to note that it is $k_c\ell_0$ instead of $\ell_0$ that should be understood as the de Sitter horizon size.
\begin{figure*}[t]
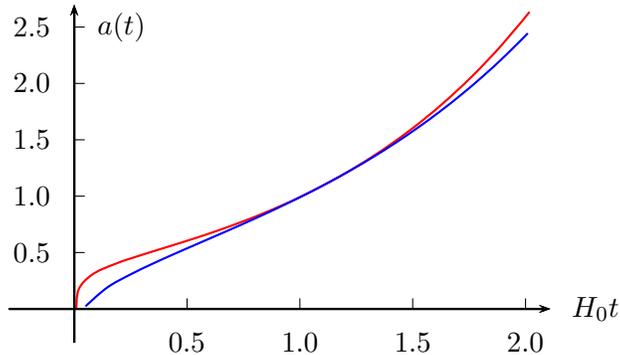

\setcaptionwidth{0.85\textwidth}
\begin{center}
\atCosmo{3}{1.5}
\caption{The evolution of scale factors in (3+1)-D statistic/gravity duality universe (red) and that of $\Lambda$CDM cosmology (blue) with $\Omega_m=0.3$, $\Omega_\Lambda=0.7$. We choose the parameter $\ell_0$ in such a way that $a(\tilde{t})=1$ as $H_0\tilde{t}=1$.}
\label{figEvolCosmo}
\end{center}
\end{figure*}
While at points far away from the infra-red fixed point, things have two sides (in the $n=1$ models, the scale factor of the dual universe will have the same asymptotics as the following first case \eqref{earlyAsymptoticB} of the $n\geqslant2$ models)
\begin{itemize}
\item if the renormalization group flow starts from (the high temperature side of Ising model) $k<k_c$, then $k\approx\frac{1}{2}(\ln\xi)^\frac{1}{2}$ as $\xi\rightarrow1^+$ and $\tilde{t}\rightarrow 0^+$, define $\frac{1}{3}\frac{\tilde{t}^{3/2}}{\ell_0^{1/2}}\equiv t$, the dual geometry will asymptote to
\beq{}
ds^2=-dt^2+a^2(t)d\vec{\bfx}^2,~
a(t)=\frac{\tilde{t}^{1/2}}{2\ell_0^{1/2}}e^{\tilde{t}/\ell_0}
=\frac{1}{2}\big(\frac{3t}{\ell_0}\big)^\frac{1}{3}e^{(3t/\ell_0)^{2/3}}
\label{earlyAsymptoticB}
\eeq
Since the evolving trend of both $\tilde{t}$ and $t$ is from $0$ to $\infty$, this kind of universe starts from a power law expanding phase, which in standard cosmologies corresponds to evolutions driven by ``super-radiations'' of equation of state $w\equiv p/\rho=+1$.
\item if the flow starts from (the low temperature side of Ising model) $k_s>k_c$, then $k\approx k_s\xi^{-n+1}=k_se^{-(n-1)t/\ell_0}$ as $\xi\rightarrow1^+$ and $\tilde{t}\rightarrow0^+$. By defining $\frac{k_s\ell_0}{n-1}e^{-(n-1)\tilde{t}/\ell_0}\equiv t$, the dual geometry can be written asymptotically as
\beq{}
ds^2=(-dt^2+a^2(t)d\vec{\bfx}^2),~a(t)=k_s^2e^{-(n-2)t/\ell_0}=k_s^2\big[\frac{(n-1)t}{k_s\ell_0}\big]^{\frac{n-2}{n-1}}
\label{earlyAsymptoticA}
\eeq
In this case the evolving trend of $\tilde{t}$ is from $0$ to $\infty$ while that of $t$ is from $\frac{k_s\ell_0}{n-1}$ to $0$, so the corresponding universe starts from a power law contracting phase.
\end{itemize}
Obviously, the first case is more close to the practical universe. Complementing to these limit analysis, we give in figure \ref{figEvolCosmo} exact numerical behaviors of $a(t)$ in both the statistic/gravity dual universe and that of standard $\Lambda$CDM universe. From the figure we easily see that this two model have rather similar evolution features. Of course, we could not expect them to coincide exactly due to the arbitraryness of our choosing of \eqref{betaUniverse} to control the running of $k$ in statistic sides. However, it'll be very exciting to find statistic models whose renormalization group equation happens to give the desired evolution of the practical universe.

Although our putting a minus in the beta function \eqref{betaUniverse} relative to the faithful Ising model in the above analysis is very artificial, the idea of identifying time as an emergent concept from lower dimensional ESS or EFT is very reasonable (because this only means that we identify the time flow as the inverse RGF when the fixed point of statistic models is unstable) and may project remarkable highlights to the solving of cosmological constant problem. Most importantly, this idea implies that, the accelerational expansion of the practical $(3+1)$-D cosmology is related only with the vaccum energy of some 3-D ESS or EFT; while the vaccum energy of $(3+1)$-D quantum field theory will affect expansion of universes only in $(4+1)$-D or $(3+2)$-D worlds.

\section{Evidence developing}
The matching of symmetry properties is not the only evidence for our ds/CFTe or statistic/gravity duality. We develope in this section two addition piece of evidences for this duality. The first is the matching of 2-point correlation functions calculated from this duality. The second is about explanations for the horizon entropy of the deSitter geometry.

As is well known, on the infra-red fixed point the statistic (e.g. Ising) models  become conformal field systems. In such systems the two point correlation function $\langle\mathcal{O}(\vec{\bfx})\mathcal{O}(\vec{\bfx}')\rangle$ of massless probe degrees of freedom is power-law decreasing as distance increases. While according to the statistic/gravity duality (in this case being just the dS/CFTe),  the corresponding correlation can be calculated through the generatating function
\beq{}
Z_{CFTe}[\hat{\phi}]\equiv\int\!\!\mathcal{D}X\,e^{iS_{CFTe}[X,\hat{\phi}]}
=Z_{ds}[\hat{\phi}]
\xLongrightarrow[clss.grav.]{saddle}e^{iS[g_{ds},\phi]^{o.s.}_{\hat{\phi}\,b.c.}}
\eeq
where $\hat{\phi}$ on the left hand side denotes perturbations coupled with $\mathcal{O}$, while $\hat{\phi}$ on the right hand side is just the boundary value of classic probes living on the de Sitter background,
\beq{}
S[g_{ds},\phi]=\int\!dx^n\!dt\,(-\frac{1}{2}g^{\mu\nu}_{ds}\partial_{\!\mu}\phi\,\partial_{\!\nu}\phi-\frac{1}{2}m^2\phi),~g_{ds}=\mathrm{diag}\{-\frac{\ell^2}{\tau^2},\frac{\ell^2}{\tau^2},\cdots\}.
\label{actionProbefield}
\eeq
Fourier expanding $\phi(\tau,\bfx)={\displaystyle\int}\!\phi_{\bfk}(\tau)e^{i\bfk\cdot\bfx}d\bfk$, we will find that $\phi_\bfk(\tau)$ following from the action principle controlled by \eqref{actionProbefield} is nothing but the bessel function of Hankel type up to a $(k\tau)^\frac{n}{2}$ power factor
\beq{}
\phi_\bfk(\tau)=(k\tau)^\frac{n}{2}[C_1H^{(1)}_\nu(k\tau)+C_2H^{(2)}_\nu(k\tau)],~\nu=(n^2/4-m^2\ell^2)^\frac{1}{2}
\eeq
Comparing with similar calculations in AdS/CFT(minkowski) correspondences \cite{AdSCFTreview}, things here are different only by $kz\equiv\sqrt{-k_0^2+\bfk^2}\cdot z$ replaced with $\sqrt{\bfk^2}\cdot\tau$ and $I_\nu(kz), K_\nu(kz)$ replaced with $H_\nu^{(1,2)}(k\tau)$. Further calcuations are also parallel with those in AdS/CFTs, except that we replace the normalizable condtion in the $z\rightarrow\infty$ limit of AdS case with the infalling(to infinity) condtion in the $\tau\rightarrow\infty$ limit so that $\phi(\tau,\bfx)\xrightarrow{\tau\rightarrow\infty} e^{-ik\tau+ikx}$. The final result is
\beq{}
\langle\mathcal{O}(\bfx)\mathcal{O}(\bfx')\rangle=\lim_{\tau\rightarrow0}\int\frac{\delta^2S[g_{ds},\hat{\phi}_\mathbf{q}]}{\delta\hat{\phi}_\bfk(\tau)\delta\hat{\phi}_{\bfk'}(\tau)}e^{-i\bfk\cdot\bfx-i\bfk'\cdot\bfx'}d^{n}\!\bfk d^n\!\bfk'=\delta(\bfx-\bfx')+\frac{c}{|\bfx-\bfx'|^{n+2\nu}}
\eeq
So after removing the point contact term $\delta(\bfx-\bfx')$, this is just a power law decreasing correlation as expected.

Our second evidence for the statistic/gravity duality is about the horizon entropy of de Sitter geometry. For very long times, peoples know that inertial observers in deSitter space can detect  thermal radiation of temperature $T$ and the corresponding entropy $S$,
\beq{}
T=\frac{1}{2\pi\ell}
,~
S_{dS_{n+1}}=\frac{\ell^{n-1}}{4G}(n\geqslant2),~0(n=1)
\eeq
 But the microscopic origin of this $T$ and $S$ is an outstanding mystery, see \cite{dSentropyMS1998, linFL1999, hawking2001} for explanations based on Strominger's dS/CFT duality. However, if our statistic/gravity duality is true, then $T$ is nothing but the crtitical temperature (zero for 1D) of the dual ising model
 \beq{}
 T=\frac{1}{2\pi k_c\ell_0}=\frac{k_{_B}T_c}{2\pi \mu\ell_0},
 \eeq 
 While about the entropy $S$, we conjecture it is just the single-site entropy of the model on   
 conformal fixed points, which can be caclulated exactly at least for $n=1,2,3$ (the $n=3$ expression is from \cite{ising3Dexact}) case,
 \beq{}
 Z(n=1)=(e^{k}+e^{-k})^\mathcal{N}\xrightarrow{k\rightarrow k_c=\infty}e^{k_c\mathcal{N}},
 \eeq
 \beq{}
 \frac{\ln Z(n=2)}{\mathcal{N}}=\ln2+\frac{1}{2}\iint_{-\pi}^\pi\frac{dq_xdq_y}{(2\pi)^2}\ln\big[
 \mathrm{ch}^22k- \mathrm{sh}2k(\cos q_x+\cos q_y)\big]
 \eeq
 \beq{}
 \frac{\ln Z(n=3)}{\mathcal{N}}=\ln2+\frac{1}{2}\iiiint_{-\pi}^\pi\!\!\!\frac{dq_xdq_ydq_zdq_w}{(2\pi)^4}\ln\big[
 \mathrm{ch}2k\,\ch6k-\sh2k\cos{q_w}-\frac{1}{3}\mathrm{sh}6k(\cos q_x+\cos q_y+\cos q_z)\big]
 \eeq
 \beq{}
 \frac{S}{\mathcal{N}}=\frac{k_{_B}}{\mathcal{N}}(\ln{Z}-\beta\frac{\partial}{\partial\beta}\ln Z)
 =\frac{k_{_B}}{\mathcal{N}}(\ln{Z}-k\frac{\partial}{\partial k}\ln Z)
 \eeq
 Obviously, for the $n=1$ case, $S^{1site}_{1D,ising}=0$ as expected. But the $n=2$ and $3$ case contain subleties. In such cases, calculations according to formulas above can only tell us that 
$S^{1site}_{nD,ising}$ equates finite numbers. They cannot tell us the number is proportional to the corresponding $k_c^{n-1}$. But we note that
\beq{}
\frac{S^{1site}_{3,ising}}{(\sqrt{8}\pi k_3)^2}|_{k=k_{3Dc}}=1.00545\times\frac{S^{1site}_{2,ising}}{(\sqrt{8}\pi k_{2})^1}|_{k=k_{2C}}.
\eeq
This may be a hint pointing to the area law of deSitter space horizon entropies, but based on a completely statistical explanation.

\section{Conclusion}
Basing on the so called statistic/gravity duality, we propose in this paper the idea of identifying the time flow of $(n+1)$D universe with the renormalization group flow (inverse if necessary) of some $n$D statistic models at equilibriums, thus making time emerge naturally from lower dimensional physic systems without time concepts. For $3$D Ising model, the dual universe with emergent time flow has qualitatively the same evolution history as the real world. By the 2 point correlation function's calculation and the de Sitter entropy's statistic explanation, we provide evidences for the validity of our basis ---  statistic/gravity duality. 

The idea of identifying RGF with the cosmic time flow may project remarkable highlights on the solving of cosmological constant problem. Since if this idea is the case, then the acceleration of $(3+1)$D universe has relevance only with properties of some $3$D equilibrium statistic model or euclidean field theories; while the vacuum energy of $(3+1)$D quantum field theories could affect dynamics only of some $(3+2)$ or $(4+1)$D universe. Of course, to establish this idea and its basis --- statistic/gravity duality, we have long way to go.

\section*{Acknowledgements}
This work is supported by Beijing Municipal Natural Science Foundation, Grant. No. Z2006015201001.

\end{CJK}
\end{document}